\documentclass[galaxies,article,accept,pdftex,moreauthors]{Definitions/mdpi} 

\firstpage{1} 
\pubvolume{1}
\issuenum{1}
\articlenumber{0}
\pubyear{2024}
\copyrightyear{2024}
\datereceived{ } 
\daterevised{ } % Comment out if no revised date
\dateaccepted{ } 
\datepublished{ } 
\hreflink{https://doi.org/} % If needed use \linebreak
\usepackage{color}
\Title{Implications of the intriguing constant inner mass surface density observed in dark matter halos}

\TitleCitation{Constant mass surface density in dark matter halos}

\Author{Jorge S\'anchez Almeida$^{1,2}$*\orcidA{0000-0003-1123-6003}}

\AuthorCitation{S\'anchez Almeida, J.}
\address{%
$^{1}$ \quad  Instituto de Astrof\'\i sica de Canarias, La Laguna, Tenerife, E-38200, Spain\\
$^{2}$ \quad Departamento de Astrof\'\i sica, Universidad de La Laguna, Tenerife, Spain}

\corres{Correspondence: jos@iac.es}

\abstract{It is known for long that the observed mass surface density of cored dark matter (DM) halos is approximately constant, independently of the galaxy mass (i.e., $\rho_c r_c\simeq {\rm constant}$, with $\rho_c$ and $r_c$ the central volume density and the radius of the core, respectively). Here we review the evidence supporting  this empirical fact as well as its theoretical interpretation. It seems to be an emergent law resulting from the concentration-halo mass relation predicted by the current cosmological model, where the DM is made of collisionless cold DM particles (CDM). 
We argue that the prediction $\rho_c r_c\simeq {\rm constant}$ is not specific to this particular model of DM but  holds for any other DM model (e.g., self-interacting) or process  (e.g., stellar or AGN feedback) that redistributes the DM within halos conserving its CDM mass.   
In addition, the fact that $\rho_c r_c\simeq {\rm constant}$  is shown to allow the estimate of the core DM mass and baryon fraction from stellar photometry alone, particularly useful when the observationally-expensive conventional spectroscopic techniques are unfeasible. 
}

\begin{document}

%%%%%%%%%%%%%%%%%%%%%%%%%%%%%%%%%%%%%%%%%%
\section{Introduction}\label{sec:introduction}

The shape of the dark matter (DM) halos hosting galaxies can be inferred from rotation curves or other kinematical measurements \cite[e.g.,][]{1996MNRAS.281...27P,2015AJ....149..180O,2019A&ARv..27....2S}. The resulting DM radial profiles often show an inner plateau  or {\em core} characterized by a central mass density $\rho_c$ and a core radius $r_c$ which combined happen to yield a surface density approximately constant, 
\begin{equation}
  \rho_c r_c\simeq {\rm constant},
\label{eq:maineq}
\end{equation}
a property observed to hold in a wide range of halo masses $M_h$, between $10^9$ and $10^{12}\,M_\odot$
\cite{1995ApJ...447L..25B,2000ApJ...537L...9S,2008MNRAS.383..297S,2009MNRAS.397.1169D,2012MNRAS.420.2034S,2014MNRAS.445.3512S,2015ApJ...808..158B,2016ApJ...817...84K,2019MNRAS.490.5451D} (actual values and details will be given in Sect.~\ref{sec:observations} and Appendix~\ref{app:literature}).  Originally, it was a rather surprising result \cite{1995ApJ...447L..25B} but nowadays it is interpreted in the literature as an emergent law caused by the well known relation between halo mass and concentration arising in collisionless cold dark matter (CDM) numerical simulations \cite[][]{2016JCAP...03..009L,2020ApJ...904..161B,2024PASJ...76.1026K}. In CDM-only simulations, the CDM halos do not have cores. They follow the canonical NFW profiles \cite{1997ApJ...490..493N} or the Einasto profiles \cite{2020Natur.585...39W}, with a pronounced inner cusp where the density grows continuously toward the center of the halo. Thus, an additional physical process must operate to transform the cuspy CDM halos into cored halos, conserving the original DM mass. This transformation is usually assumed to be driven by baryon processes like star-formation feedback, AGN feedback, or galaxy mergers, which shuffle around the baryonic mass, thus changing the overall gravitational potential and affecting the distribution of CDM. CDM cores appear in model galaxies formed in full hydrodynamical cosmological numerical simulations \cite[e.g.,][]{2010Natur.463..203G,2014Natur.506..171P,2020MNRAS.497.2393L}. Thus, Eq.~(\ref{eq:maineq}) is often regarded as a support for CDM \cite[][and references therein]{2024PASJ...76.1026K}. However, the formation of  cores in DM halos can be driven by any physical processes that thermalizes the DM distribution \cite{1993PhLA..174..384P,2020A&A...642L..14S}. They will also render Eq.~(\ref{eq:maineq}), provided the process just redistributes the available mass, not changing much the relation between halo mass and concentration set by the cosmological initial conditions (Sect.~\ref{sec:what_sets}).

The purpose of this work is to review the observational evidence for Eq.~(\ref{eq:maineq}) as well as the theory behind it. The interpretation can be pinned down to the relation between the mass of a DM halo and its age of formation (Sect.~\ref{sec:what_sets}), which is set by cosmology and to a lesser extent by details on the nature of DM. As a spin-off, we demonstrate how Eq.~(\ref{eq:maineq}) can be used to estimate the mass in the DM halo of a galaxy based solely on the distribution of its stars. The approach is based on the fact that  dwarf galaxies also tend to show a central plateau or core in the {\em stellar} distribution \cite[e.g.,][]{2021ApJ...922..267C,2024A&A...681A..15M}. The radii of the stellar and the DM cores are expected to scale with each other \cite{2024ApJ...973L..15S,SanchezAlmeida24}. We worked out the relation between the core radius of the stellar distribution and the DM mass.

The paper is organized as follows:
Section~\ref{sec:observations} collects observational evidence for Eq.~(\ref{eq:maineq}). 
Section~\ref{sec:theory} works out the explanation of Eq.~(\ref{eq:maineq}) within CDM.
Section~\ref{sec:theory_observations} compares the observations in Sect.~\ref{sec:observations} with the theory in Sect.~\ref{sec:theory}.
Based on Eq.~(\ref{sec:theory_observations}),  Sect.~\ref{sec:discussion} writes down a semi-empirical relation between the stellar core radius and DM halo mass. It also shows that the stellar mass surface density is a proxy for the baryon fraction in the center of a galaxy. Ready to use relations are given in Eqs.~(\ref{eq:breakthrough1b}) and (\ref{eq:breakthrough2b}). 
Section~\ref{sec:conclusions} summarizes the main conclusions in the work.

%
%%%%%
%
%
\begin{figure}
\centering 
\includegraphics[width=0.8\linewidth]{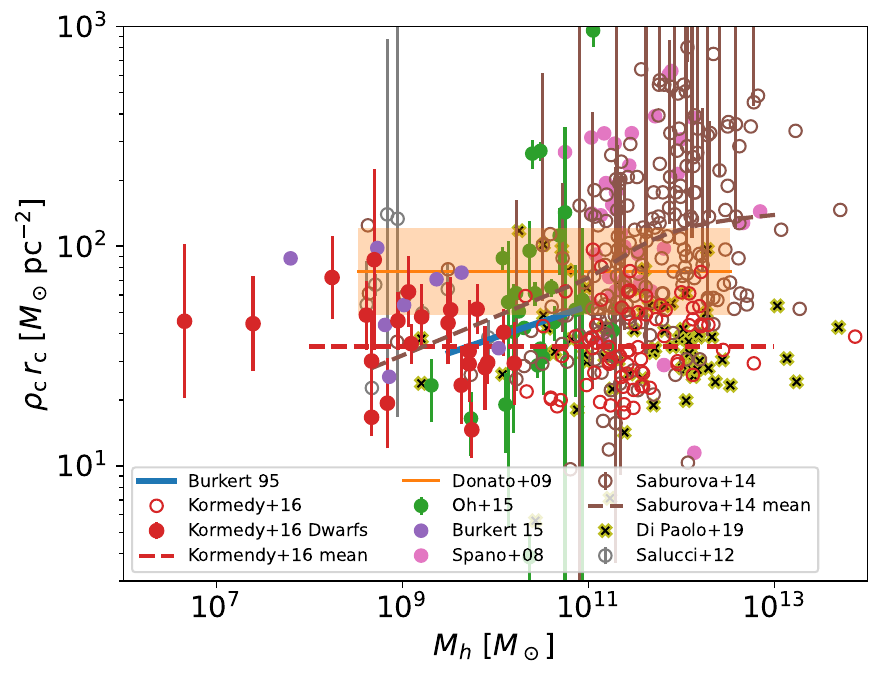}
\caption{
Compilation of values of $\rho_c r_c$ from the literature as a function of the DM halo mass of the galaxy ($M_h$). Details on the references and the processing are given in Appendix~\ref{app:literature}. A version of this figure but showing the same  eight orders of magnitude range for abscissae and ordinates is shown in Fig.~\ref{fig:core_relation6_a}. 
References: Burkert~95\,\cite{1995ApJ...447L..25B}, Kormendy+16\,\cite{2016ApJ...817...84K}, Donato+09\,\cite{2009MNRAS.397.1169D}, Oh+15\,\cite{2015AJ....149..180O}, Burkert~15\,\cite{2015ApJ...808..158B}, Spano+08\,\cite{2008MNRAS.383..297S}, Saburova+14\,\cite{2014MNRAS.445.3512S}, Di~Paolo+19\,\cite{2019MNRAS.490.5451D}, and Salucci+12\,\cite{2012MNRAS.420.2034S}.
The inset gives a color and symbol code which is the same used in Figs.~\ref{fig:core_relation6f}, \ref{fig:core_relation6b}, and \ref{fig:core_relation6d}. 
}
\label{fig:core_relation6}
\end{figure}
\begin{figure}
\centering 
\includegraphics[width=0.6\linewidth]{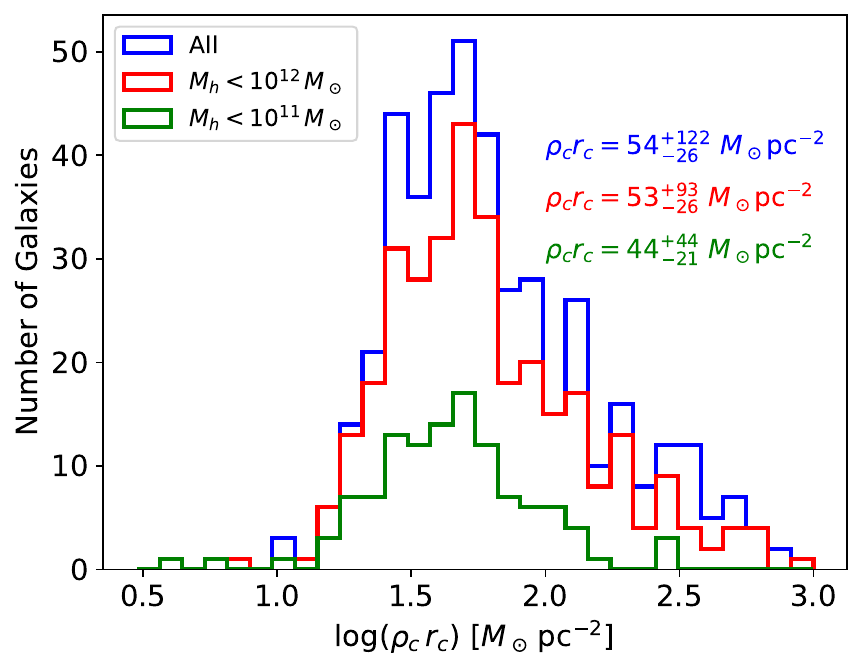}
\caption{
Histograms with the distribution of $\rho_c r_c$ represented in Fig.~\ref{fig:core_relation6} and detailed in Appendix~\ref{app:literature}.  We show three different selections: all galaxies (the blue line), galaxies with halo masses $M_h< 10^{12}\,M_\odot$ (the red line), and galaxies with $M_h< 10^{11}\,M_\odot$ (the green line). The last one is representative of dwarf galaxies. The inset gives the median of each distribution, as well as the range between percentiles 15.9\,\% and 84.1\,\% (i.e., median $\pm 1\,$sigma).   
}
\label{fig:core_relation6g}
\end{figure}
\section{Observations supporting Eq.~(\ref{eq:maineq})}\label{sec:observations}
As we point out in Sect.~\ref{sec:introduction}, the product $\rho_cr_c$ is approximately constant over a large range in galaxy mass.  To emphasize the existing evidence, we have compiled a number of relations between $\rho_c r_c$ and $M_h$ from the literature.  They are based on uneven measurements prone to bias, including the determination of the DM halo mass of a galaxy and the definition of core radius. However, the conclusion is clear, with the different independent determinations agreeing within error bars. The result of the compilation is shown in Figs.~\ref{fig:core_relation6}\,--\,\ref{fig:core_relation6d}. Details of how the individual works were interpreted to construct the figures are given in Appendix~\ref{app:literature}. In particular, here and throughout the paper, we assume the core radius to be the radius where the density drops to half the central value,
\begin{equation}
  \rho(r_c)= \rho_c/2,
  \label{eq:core_def}
\end{equation}
with $\rho_c=\rho(0)$. This definition is not universally used and so often the radii quoted in the original reference have to be transformed to our definition, as detailed in Appendix~\ref{app:literature}.

\begin{figure}
\centering 
\includegraphics[width=0.7\linewidth]{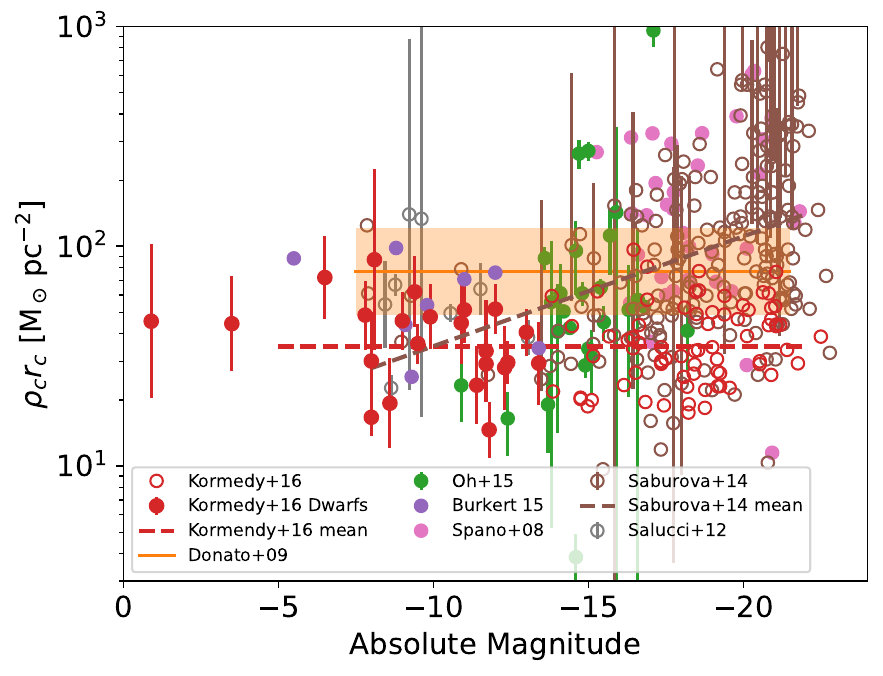}
\caption{
Central DM surface density, $\rho_c r_c$, as a function of the absolute magnitude of the galaxy, which is the observable employed to estimate the halo masses represented in Fig.~\ref{fig:core_relation6}. The absolute magnitude is $M_B$ or $M_V$ depending on the galaxy. The inset gives the color and symbol code, which is the same employed in  Figs.~\ref{fig:core_relation6}, \ref{fig:core_relation6b}, and \ref{fig:core_relation6d}. 
}
\label{fig:core_relation6f}
\end{figure}

\begin{figure}
\centering 
\includegraphics[width=0.6\linewidth]{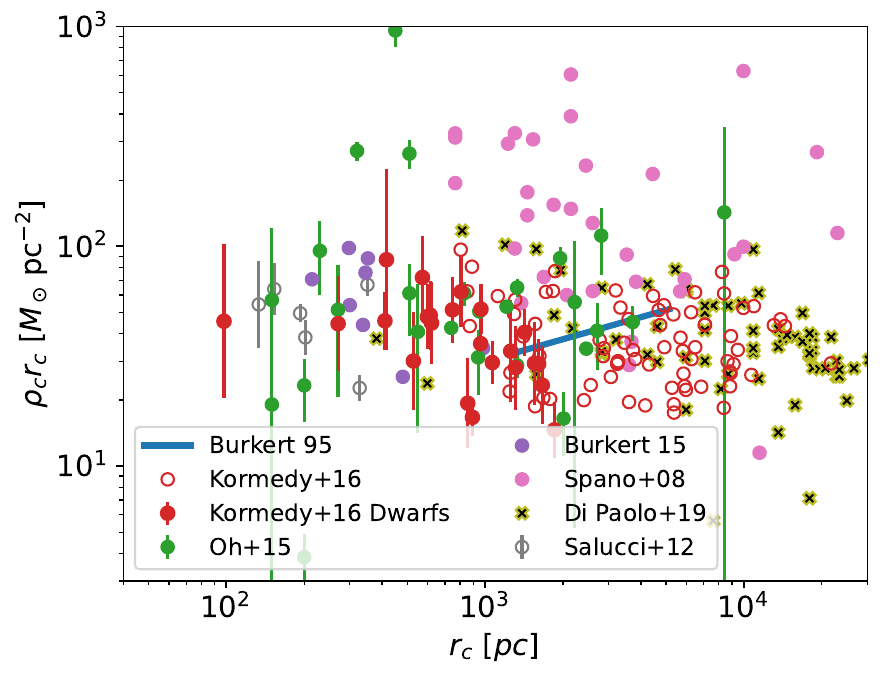}
\includegraphics[width=0.6\linewidth]{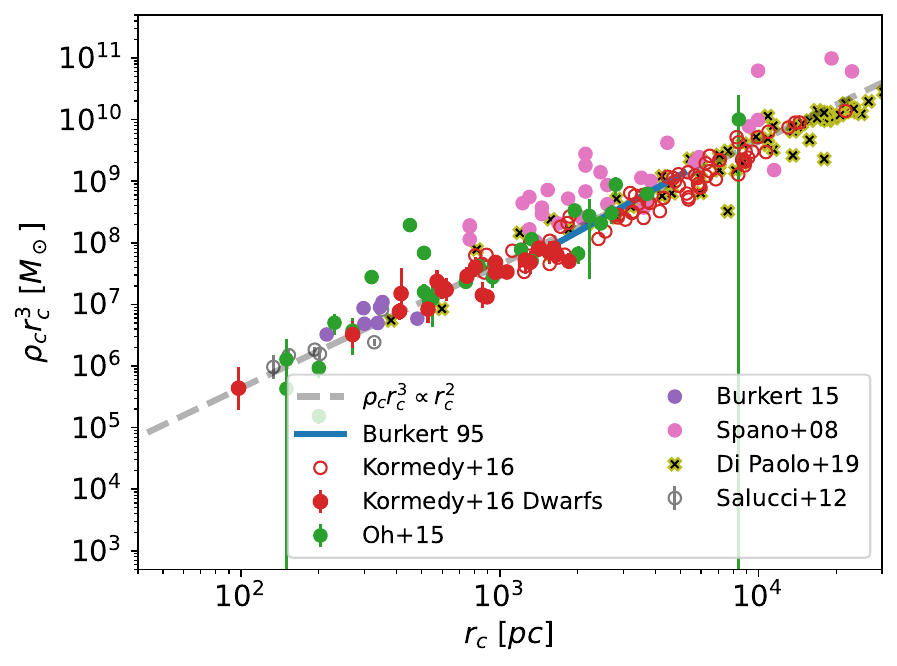}
\caption{Observed  $\rho_cr_c$ versus  $r_c$ (top panel) and  $\rho_cr_c^3$  versus  $r_c$ (bottom panel). Note that the latter gives the DM mass in the core and scales as $r_c^2$ following Eq.~(\ref{eq:obs_rhor}), which is represented by the gray dashed line.  These relations do not depend on the total DM halo mass and can be used to test theoretical explanations bypassing uncertainties in $M_h$. The insets give the color and symbol code, used also in  Figs.~\ref{fig:core_relation6}, \ref{fig:core_relation6f}, and \ref{fig:core_relation6d}. 
}
\label{fig:core_relation6b}
\end{figure}

\begin{figure}
\centering 
\includegraphics[width=0.7\linewidth]{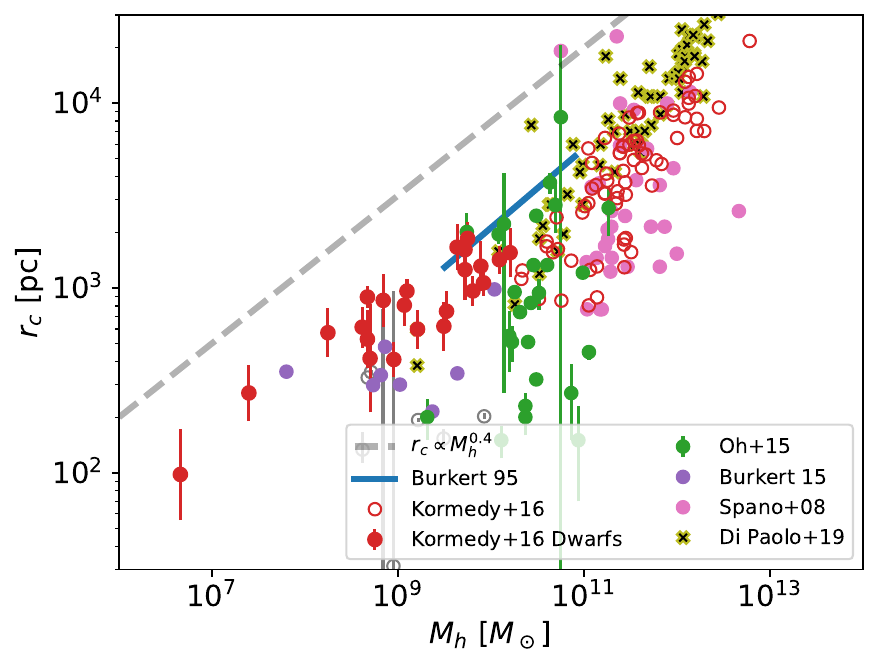}
\includegraphics[width=0.7\linewidth]{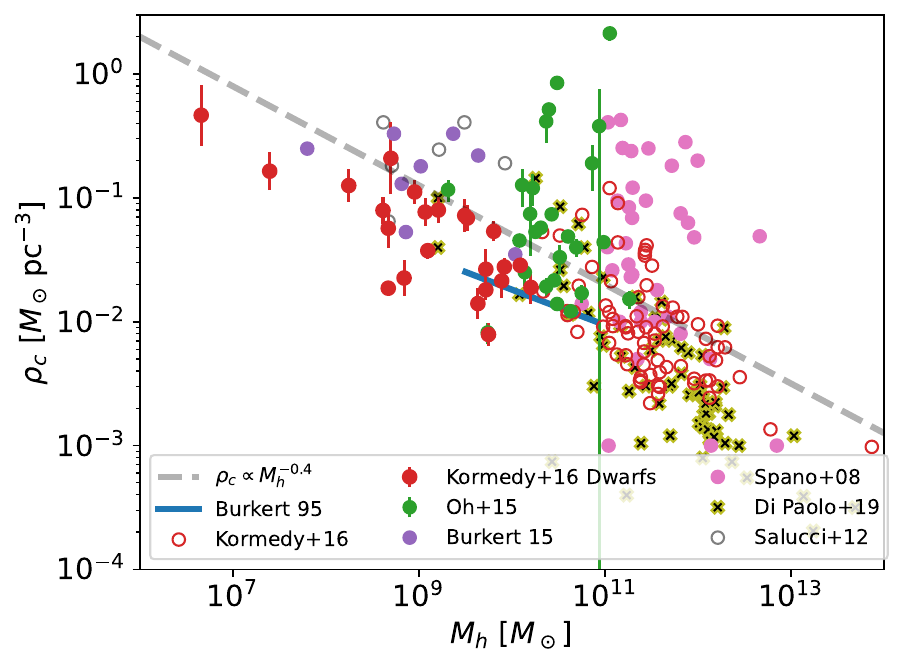}
\caption{
  Top panel: core radius $r_c$ versus DM halo mass $M_h$. The dashed line is a power law with exponent $+0.4$ and has been included to guide the eye.   Bottom panel: central DM density $\rho_c$ versus DM halo mass.  This time the dashed line is a power law with exponent $-0.4$. The insets give the color and symbol code, which is the same used in  Figs.~\ref{fig:core_relation6}, \ref{fig:core_relation6f}, and \ref{fig:core_relation6b}.  
}
\label{fig:core_relation6d}
\end{figure}

Figure~\ref{fig:core_relation6} gives the scatter plot of $\rho_cr_c$ versus $M_h$. The extreme values are likely unreliable  but it is clear that  the product $\rho_cr_c$ tends to be constant, at least for $M_h < 10^{11} M_\odot$.\footnote{The increased scatter for $M_h > 10^{11} M_\odot$ may be artificially caused by the challenges of disentangling the baryonic contribution from the overall potential, which must be subtracted from the observables to derive the dark matter (DM) distribution.} This fact is better appreciated in Fig.~\ref{fig:core_relation6_a}, which is identical to Fig.~\ref{fig:core_relation6}  but with the vertical axis spanning  the same eight orders of magnitude of the horizontal axis corresponding to the DM halo masses. Histograms with the values of $\rho_cr_c$ in  Fig.~\ref{fig:core_relation6} are shown in Fig.~\ref{fig:core_relation6g}. They include all the observed values (the blue line), when  $M_h < 10^{12} M_\odot$ (the red line), and when  $M_h < 10^{11} M_\odot$ (the green line). An inset in the figure also gives the median and the 1-sigma percentiles of the distributions (i.e., 50\,\%, 15.9\,\% and 84.1\,\%) which correspond to
\begin{equation}
  \rho_c r_c = 44^{+44}_{-21}\, M_\odot\,{\rm  pc}^{-2},
  \label{eq:obs_rhor}
\end{equation}
when $M_h < 10^{11} M_\odot$, a limit representative of dwarf galaxies. We note that the used $r_c$, as set by Eq.~(\ref{eq:core_def}), is typically a factor of two smaller than the core radii commonly defined in the literature\footnote{Rather than being the radius where the density drops to half the central value (Eq.~[\ref{eq:core_def}]), it is some characteristic radius defining the analytic cored density profile used in each specific paper. For example, $b$ when the Schuster-Plummer profile in Eq.~(\ref{eq:plummer}) is used.}.  Thus, the surface density in Eq.~(\ref{eq:obs_rhor}) is fully consistent with a value around $100\,M_\odot\,{\rm  pc}^{-2}$ often quoted in the literature  (see, e.g., \cite{2000ApJ...537L...9S,2020ApJ...904..161B}). 
As we explain in Appendix~\ref{app:literature}, the estimate of $M_h$ used in Fig.~\ref{fig:core_relation6} relies on the observed absolute magnitude of the galaxies, assuming a mass-to-light ratio and a relation between stellar mass and DM halo mass as inferred from abundance matching \cite{2013ApJ...770...57B}. However,  the trend for $\rho_c r_c$ to become constant in dwarf galaxies is already present in the original data; see Fig.~\ref{fig:core_relation6f}, where the abscissa are given by the measured absolute magnitude of the galaxy. 
Figure~\ref{fig:core_relation6b} gives  $\rho_cr_c$ (top panel) and $\rho_cr_c^3$ (bottom panel) versus  $r_c$. Note that the latter gives the DM mass in the core and it scales as $r_c^2$ following Eq.~(\ref{eq:obs_rhor}), which is represented in the figure by the gray dashed line. These relations are  independent of the uncertainties in $M_h$.

Figure~\ref{fig:core_relation6d}   gives the relation  of  $r_c$ with $M_h$  (top panel)  and $\rho_c$ with $M_h$ (bottom panel). The correlation happens to be very clear in both cases. The larger the mass, the larger the radius and the smaller the density. In order to guide the eye, the figure includes power laws
as $r_c\propto M_h^{0.4}$ (top panel) and $\rho_c\propto M_h^{-0.4}$ (bottom panel), which approximately describe the observed trends. Note that combined, these power laws render Eq.~(\ref{eq:maineq}). 

\section{Theory: cores resulting from redistributing collisionless cold dark matter halos}\label{sec:theory}

\begin{figure}
\centering 
\includegraphics[width=0.75\linewidth]{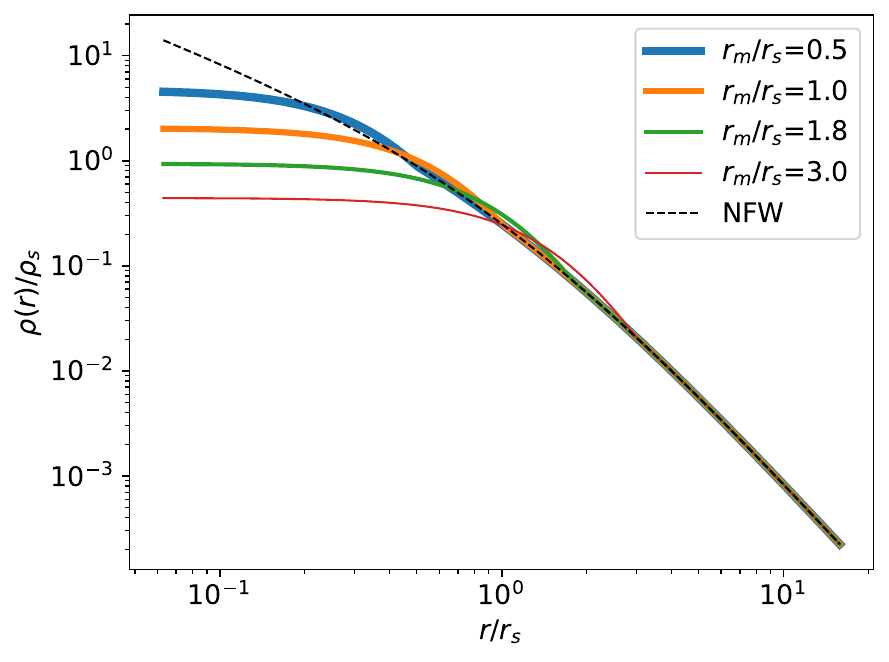}
\caption{
Piecewise density profiles with an inner core ($m=5$ polytrope; $\rho_5$ in Eq.~[\ref{eq:plummer}]) and an outer NFW profile ($\rho_{\rm NFW}$; Eq.~[\ref{eq:nfw}]). The two pieces coincide at the matching radius $r_m$, $\rho_5(r_m)=\rho_{\rm NFW}(r_m)$, and the total mass is the total mass of $\rho_{\rm NFW}(r)$ (Eq.~[\ref{eq:massc}]). The full NFW profile is shown as a black dashed line whereas profiles for different matching radii are shown with different colors as indicated in the inset.
}
\label{fig:core_relation4}
\end{figure}

If the DM was collisionless CDM and if there were no baryons,  then the distribution of DM within each halo would approximately  follow the iconic NFW profile \cite{1997ApJ...490..493N},
\begin{equation}
  \rho_{\rm NFW}(r) = \frac{\rho_s}{(r/r_s)\,(1+r/r_s)^2},
\label{eq:nfw}
\end{equation}
describing the variation with radius $r$ of the DM volume density $\rho_{\rm NFW}(r)$. The parameters $r_s$ and $\rho_s$ stand for a scaling radius and a scaling density, respectively.  The mass available to form any DM halo today is provided by the initial conditions set by cosmology (see Sect.~\ref{sec:what_sets}). It would be the same independently of whether a physical process redistributes this mass in a different mass density profile. Probably, the most general such process is the thermalization the DM distribution. In this case, one expects the formation of a core with a generic polytropic shape, characteristic of self-gravitating systems reaching thermodynamic equilibrium \cite{1993PhLA..174..384P,2020A&A...642L..14S,2022Univ....8..214S}. For analytic simplicity, we assume the $m=5$ polytrope (best known as Schuster-Plummer profile), but the core of all polytropes has virtually the same shape \cite[e.g.,][]{2022Univ....8..214S}. In this case,
\begin{equation}
\rho_5(r) =\frac{\rho_c}{\left[1+(r/b)^2\right]^{5/2}},
\label{eq:plummer}
\end{equation}
with $\rho_c$ the central density and $b$ a length scale setting the core radius defined as in Eq.~(\ref{eq:maineq}),
\begin{equation}
  r_c= b\times\sqrt{2^{2/5}-1}\simeq b\times 0.56525\dots.
  \label{eq:silly}
\end{equation}
Thus, the new density profile resulting from the core formation is a piecewise function defined as Eq.~(\ref{eq:plummer}) in the core, Eq.~(\ref{eq:nfw}) in the outskirts, and  continuous in the matching radius $r_m$,
\begin{equation}
  \rho(r)=
  \begin{cases}
\rho_5(r), & {\rm when~~} r <  r_m,\\
\rho_5(r_m)=\rho_{\rm NFW}(r_m),& {\rm when~~} r=r_m,\\
\rho_{\rm NFW}(r),& {\rm when~~} r > r_m.
\end{cases}
\label{eq:cases}
\end{equation}
In addition, to conserve mass,
\begin{equation}
  \int_0^\infty \rho(r)\,r^2\,dr = \int_0^\infty \rho_{\rm NFW}(r)\,r^2\,dr,
  \label{eq:massc}
\end{equation}
which, considering Eq.~(\ref{eq:cases}), renders
\begin{equation}
  \int_0^{r_m} \rho_5(r)\,r^2\,dr = \int_0^{r_m} \rho_{\rm NFW}(r)\,r^2\,dr.
\label{eq:massconservation}
\end{equation}
Examples of these cored DM profiles with NFW outskirts are given in Fig.~\ref{fig:core_relation4}. This kind of piecewise shape has already been used in the literature \cite[e.g.,][]{2021MNRAS.501.4610R,2021MNRAS.504.2832S}.
Equations~(\ref{eq:cases}) and (\ref{eq:massconservation}) provide a mapping between the parameters of the NFW profile ($\rho_s$ and $r_s$) and the parameters defining the core ($\rho_c$ and $b$).
The continuity at $r_m$ forces
\begin{equation}
  \frac{\rho_c}{(1+(r_m/b)^2)^{5/2}}=\frac{\rho_s}{(r_m/r_s)(1+r_m/r_s)^2},
  \label{eq:continuity}
\end{equation}
whereas mass conservation, Eq.~(\ref{eq:massconservation}), leads to
\begin{equation}
\rho_c\,r_m^3\,  \frac{1}{3\,[1+(r_m/b)^2]^{3/2}}=
\rho_s\,r_s^3\,\left[\ln(1+\frac{r_m}{r_s})-\frac{r_m/r_s}{1+r_m/r_s}\right].
\label{eq:conservation}
\end{equation}
After some algrebra, Eqs.~(\ref{eq:continuity}) and (\ref{eq:conservation}) render,
\begin{equation}
  1+\left(\frac{r_m/r_s}{b/r_s}\right)^2 =\frac{3\,(1+r_m/r_s)^2}{(r_m/r_s)^2}\,\left[\ln(1+\frac{r_m}{r_s})-\frac{r_m/r_s}{1+r_m/r_s}\right],
  \label{eq:constraint1}
\end{equation}
and
\begin{equation}
  \frac{\rho_c\,b^3}{\rho_s\,r_s^3} =\frac{3\,[1+(r_m/b)^2]^{3/2}}{(r_m/b)^3}\,\left[\ln(1+\frac{r_m}{r_s})-\frac{r_m/r_s}{1+r_m/r_s}\right].
  \label{eq:constraint2}
\end{equation}
We note that once $r_m/r_s$ is set (i.e., the radius of match in units of $r_s$; see Eq.~[\ref{eq:cases}]), Eqs.~(\ref{eq:constraint1}) and (\ref{eq:constraint2}) give the full density profile.  Equation~(\ref{eq:constraint1}) provides $b/r_s$, which can be used in Eq.~(\ref{eq:constraint2}) to compute $\rho_c/\rho_s$, and then $\rho(r)/\rho_s$. This is the procedure followed to compute the densities shown in Fig.~\ref{fig:core_relation4}.

\begin{figure}
\centering 
\includegraphics[width=0.7\linewidth]{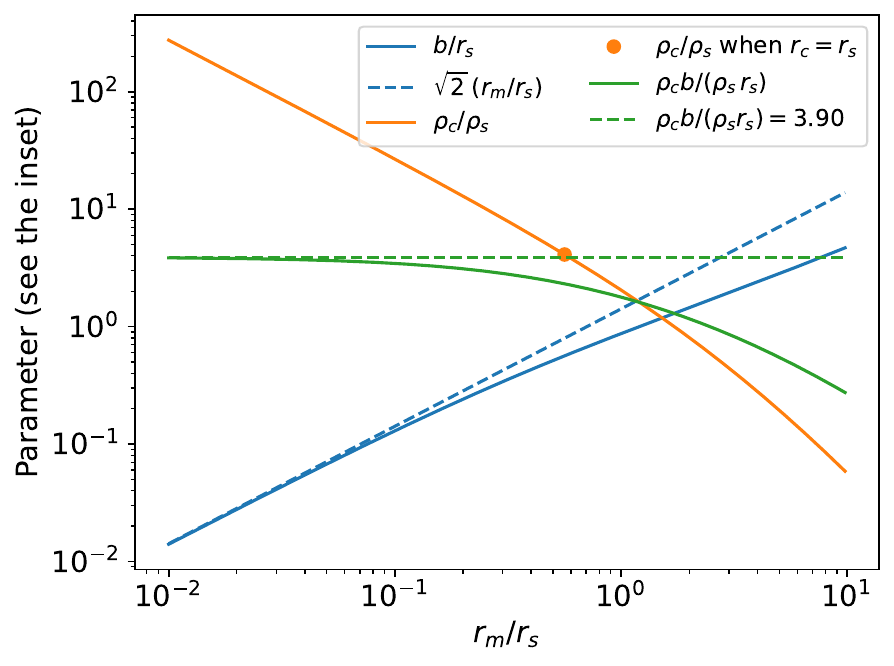}
\caption{Dependence  on $r_m/r_s$ of  $b/r_s$, $\rho_c/\rho_s$, and $\rho_c b/(\rho_s r_s)$ as given by Eqs.~(\ref{eq:constraint1}) and (\ref{eq:constraint2}). The solid lines show the actual variation whereas the dashed lines correspond to the dependence when the transition radius  $r_m \ll r_s$ (Eqs.~[\ref{eq:last}] and [\ref{eq:last2}]).  The orange symbol  points out when $r_c=r_s$, which has $r_m/r_s\simeq 0.56$ and $\rho_c/\rho_s\simeq 4.11$.
}
\label{fig:core_relation4b}
\end{figure}
Figure~\ref{fig:core_relation4b} shows the dependence on $r_m/r_s$ for $b/r_s$, $\rho_c/\rho_s$, and $\rho_c b/(\rho_s r_s)$. We note that for $r_m\lesssim r_s$,
$b\sim r_s$ and  $\rho_c b \sim \rho_s r_s$.
%
%
%\begin{equation}
%b\sim r_s
% \end{equation}
% whereas
%\begin{equation}
% \rho_c b \sim \rho_s r_s.
%\end{equation}
%
These dependences are easy to distill from the above equations in the limit $r_m\ll r_s$. In this case,
\begin{equation}
\ln(1+\frac{r_m}{r_s})-\frac{r_m/r_s}{1+r_m/r_s}\simeq \frac{(r_m/r_s)^2}{2},
\end{equation}
so that Eq.~(\ref{eq:constraint1}) renders,
\begin{equation}
  b/r_s \simeq \sqrt{2}\,(r_m/r_s).
  \label{eq:last}
\end{equation}
Similarly, Eq.~(\ref{eq:constraint2}) plus Eq.~(\ref{eq:last}) render
\begin{equation}
  \rho_c b \simeq \rho_s r_s \,(3/2)^2\sqrt{3} = \rho_s r_s\times 3.89711\dots. 
  \label{eq:last2}
\end{equation}
 When  $r_m=r_s$ (i.e., when the matching radius coincides with the characteristic radius defining the NFW profile) then things simplify even further so that,
  \begin{equation}
    \frac{\rho_c r_c}{\rho_s r_s}\simeq 1.0068\dots,
    \label{eq:magic}
\end{equation}
where we have used Eq.~(\ref{eq:silly}) to transform $b$ into $r_c$ (details in Appendix~\ref{app:appb}).

The NFW halos are given setting $\rho_s$ and $r_s$. In the context of CDM, these two variables are often replaced by the concentration\footnote{\label{this_footnote}$c=r_{200}/r_s$, with $r_{200}$ defined so that the mean enclosed density within $r_{200}$ equals 200 times the critical density $\rho_{crit}$.}  $c$ and the halo mass $M_h$, so that, 
\begin{equation}
  \rho_s = \frac{200\,c^3\,\rho_{crit}}{3\big[\ln(1+c)-c/(1+c)\big]},
  \label{eq:rho0}
\end{equation}
and,
%\begin{equation}
%M_h=4\pi\,\rho_s\,r_s^3\big[\ln(1+c)-c/(1+c)\big],
%\end{equation}
%so that
\begin{equation}
  r_s^3=\frac{3\,M_h}{800\pi\,\rho_{crit}\,c^3}.
  \label{eq:rs}
\end{equation}
The symbol $\rho_{crit}$ stands for the critical density of the Universe.  Pieced together, Eqs.~(\ref{eq:rho0}) and (\ref{eq:rs}) render the dependence of the product $\rho_s r_s$ on $c$ and $M_h$,
\begin{equation}
\rho_s r_s = \frac{10}{3}\,\left(\frac{30}{\pi}\right)^{1/3}\,\frac{\rho_{crit}^{2/3}\,c^2\,M_h^{1/3}}{\ln(1+c)-c/(1+c)},
  \label{eq:scaling1}
\end{equation}
a relation that can be found already in the literature \cite[e.g.,][]{2016JCAP...03..009L}.
\begin{figure}
\centering 
\includegraphics[width=0.6\linewidth]{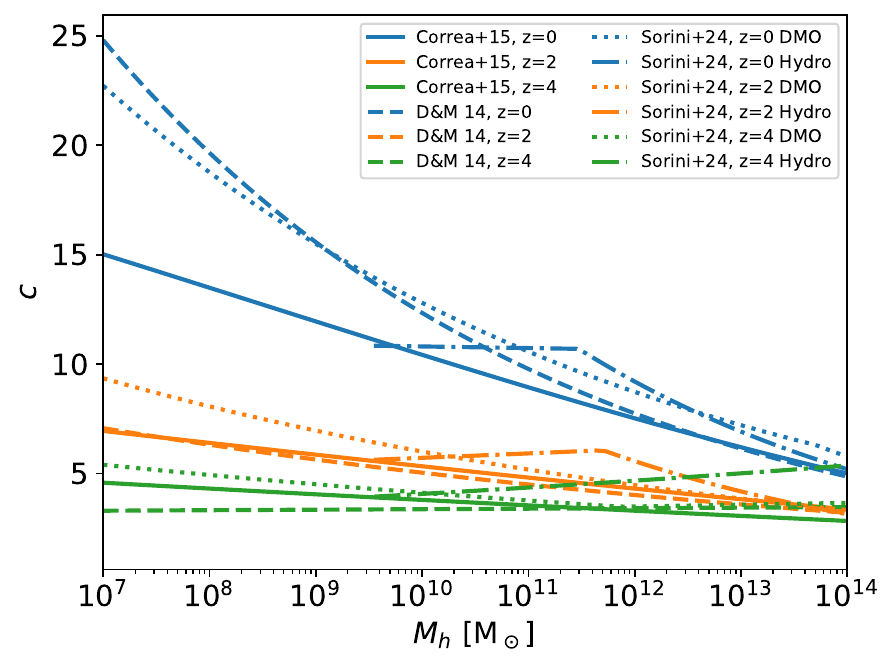}
\caption{Relation between concentration $c$ and halo mass $M_h$  inferred from various CDM only simulations. The three papers cited in the inset are  D\&M~14\,\cite{2014MNRAS.441.3359D}, Correa+15\,\cite{2015MNRAS.452.1217C}, and Sorini+24\,\cite{2024arXiv240901758S}. For reference, we also show a relation obtained when baryon feedback is self-consistently treated in the simulation (the dotted dashed lines). Different redshifts ($z$) are included with different colors, whereas the type of line encodes the actual reference (see the inset).}
\label{fig:core_relation7}
\end{figure}
The numerical simulations of CDM predict a relation between $c$ and $M_h$, which varies with redshift and is quite tight for $M_h> 10^{10}\, M_\odot$ to become looser at smaller halo mass \cite{2014MNRAS.441.3359D,2015MNRAS.452.1217C,2024arXiv240901758S}. Examples of this relation are given in Fig.~\ref{fig:core_relation7}, where we note that the range of variation of $c$ is quite moderate,  changing only by a factor of three for halos varying by seven orders of magnitude in mass, from $10^7$ to $10^{14}~M_\odot$; see the blue lines in Fig.~\ref{fig:core_relation7}. Thus, considering $c$ constant,  the dependence of $\rho_s r_s$ on halo mass predicted by  Eq.~(\ref{eq:scaling1})  is quite mild as it scales as $M_h^{1/3}$. This fact, together with the approximate equivalence given by Eqs.~(\ref{eq:silly}) and (\ref{eq:last2}), indicates that the predicted $\rho_cr_c$ is expected to vary little with halo mass,
\begin{equation}
  \rho_c r_c\propto \rho_c b\propto \rho_s r_s\propto M_h^{1/3},
  \label{eq:approx_mh}
\end{equation}
as it is indeed observed (Sect.~\ref{sec:observations}).

\begin{figure}
\begin{center} 
\includegraphics[width=0.6\linewidth]{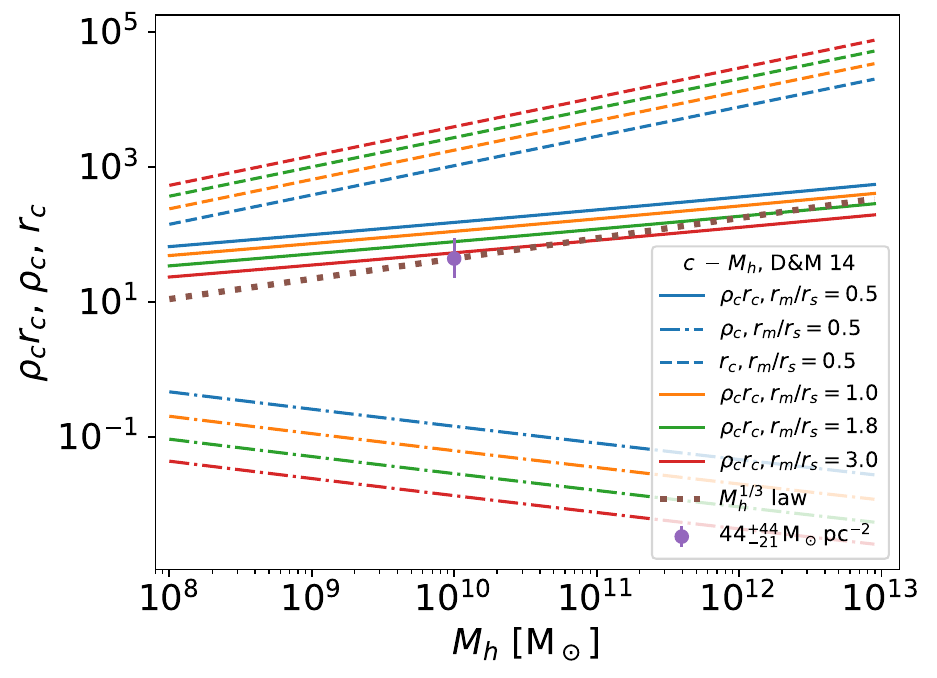}
\end{center}
\caption{
Predicted variation of the central mass surface density $\rho_c r_c$ as a function of $M_h$ for various $r_m/r_s$ assuming the $c$--$M_h$ relation at redshift zero given in \cite{2014MNRAS.441.3359D} (the solid lines). The figure also includes the variation of $r_c$ (the dashed lines) and $\rho_c$ (the dashed-dotted lines)  to emphasize how the increase of $r_c$ with increasing $M_h$ is partly balanced by the decrease of $\rho_c$ to produce a fairly constant $\rho_c r_c$. The dotted line shows the approximate dependence of $\rho_c r_c$ on $M_h$ to be expected if $c$ were constant (Eq.~[\ref{eq:approx_mh}]). This power law dependence has been anchored to the observed $\rho_c r_c$ (Eq.~[\ref{eq:obs_rhor}]) assumed to represent $M_h\sim 10^{10}~M_\odot$. The core density $\rho_c$ and core radius $r_c$ are given in units of $M_\odot\,{\rm pc}^{-3}$ and ${\rm pc}$, respectively.
}
\label{fig:core_relation2_new}
\end{figure}
The equations above yield $\rho_c r_c$ as a function of $M_h$.  The algorithm to compute it is: (1) set $r_m/r_s$, (2) get $c$ of $M_h$ from the literature (Fig.~\ref{fig:core_relation7}), (3) get $\rho_s$ and $r_s$ as a function of $M_h$ from Eqs.~(\ref{eq:rho0})  and  (\ref{eq:rs}),  (4) get $b/r_s$ of $M_h$ from $r_s$ and Eq.~(\ref{eq:constraint1}), (5) get $\rho_c/\rho_s$ of $M_h$ from $b/r_s$, $r_m/r_s$,  and  Eq.~(\ref{eq:constraint2}), (6) get $r_c/b$ from Eq.~(\ref{eq:silly}) and, finally, (7) compute
\begin{equation}
\rho_c r_c= \rho_s\times r_s \times \frac{b}{r_s}\times \frac{\rho_c}{\rho_s} \times \frac{r_c}{b}.
\end{equation}
Figure~\ref{fig:core_relation2_new} shows the predicted variation of $\rho_c r_c$ as a function of $M_h$  for various $r_m/r_s$ assuming the $c$--$M_h$ relation at redshift zero given in \cite{2014MNRAS.441.3359D} (the solid lines).  Qualitatively, the trends for other  $c$--$M_h$ relations and redshifts look the same. The figure also includes the variation of $r_c$ (the dashed lines) and $\rho_c$ (the dashed dotted lines) separately. Note how the increase of $r_c$ with $M_h$ is partly balanced by the decrease of  $\rho_c$, leaving a fairly constant $\rho_c r_c$.

%
%%%%%%%%
%
\begin{figure}
\centering 
\includegraphics[width=0.8\linewidth]{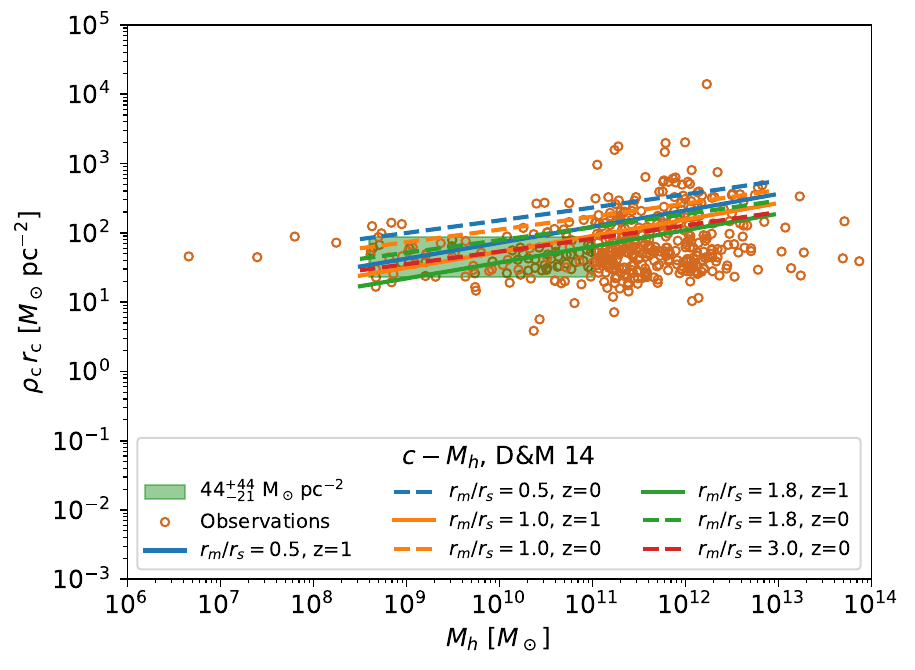}
\caption{
  Observed versus predicted $\rho_c r_c$. The observations are the same as those used in Fig.~\ref{fig:core_relation6}  except that ordinates and abscissae have been forced to span the same eight order of magnitude range. The colored lines represent the theoretical predictions, which  depend on the parameter $r_m/r_s$ and the redshift $z$ from which the $c$--$M_h$ relation was taken (see the inset). The range of  $\rho_c r_c$ values for $M_h <10^{11}\,M_\odot$ given in Eq.~(\ref{eq:obs_rhor}) is shown as the pale green region.
}
\label{fig:core_relation9}
\end{figure}
\section{Comparison between observations and theory}\label{sec:theory_observations}
Figure~\ref{fig:core_relation9} shows the observed  $\rho_c r_c$ (the symbols) compared with the prediction using the simple equations worked out in Sect.~\ref{sec:theory}, where the DM cores are assumed to result from the redistribution of mass of the CDM halos.  The observed data points in Fig.~\ref{fig:core_relation9} are those in Fig.~\ref{fig:core_relation6} but shown in a range spanning the same eight orders of magnitude variation for both $\rho_c r_c$ and $M_h$. This particular scaling evidences how constant $\rho_c r_c$ is, with the range of values in Eq.~(\ref{eq:obs_rhor}) highlighted as the pale green region. The colored lines represent the theoretical predictions and they agree  well with the observation without any fine tuning. They even reproduce a slight increase of $\rho_c r_c$ with halo mass, which is probably too large in the theoretical model, although given the observational uncertainties one should not stress this fact further.  Note that the prediction depends on the parameter $r_m/r_s$ and the redshift $z$ from which the relation $c$--$M_h$ was taken.  The best agreement with observation corresponds to $r_m/r_s$ between 1 and 2 starting off from halos at $z=0$,  and  between 0.5 and 1 starting from halos a bit earlier at $z=1$. Figure~\ref{fig:core_relation9} is based on the theoretical $c$--$M_h$ from \cite{2014MNRAS.441.3359D}, but the results are similar for the other theoretical $c$--$M_h$ analyzed in Sect.~\ref{sec:observations} and Figs.~\ref{fig:core_relation7} and \ref{fig:core_relation2_new}.

% 
%%%%%%%%%%%%%%%%%
\section{Discussion}\label{sec:discussion}

Here we analyze the implications of the fair agreement between theory and observation presented in Sect.~\ref{sec:theory_observations}. 

\subsection{What sets the $c$--$M_h$  relation?} \label{sec:what_sets}

Note that so far the answer to the question of what sets $\rho_cr_c \simeq$ constant  is {\em the existence of a  $c$--$M_h$ relation} for the DM halos produced in the $\Lambda$CDM cosmology (see Figs.~\ref{fig:core_relation7} and \ref{fig:core_relation2_new}). Thus, unless we understand in physical terms what sets the $c$--$M_h$ relation of the collisionless CDM halos,  the above explanation of why $\rho_cr_c$  is constant sounds circular.

\citet{2015MNRAS.452.1217C} describe the current understanding in detail, and give a number of relevant references.  According to this view, the relation seems to be driven by the inside-out growth of the DM halos combined with the fact that low mass halos collapse first. The build-up of all halos generally consists of an early phase of fast accretion and a late phase where the accretion slows down \cite{2003MNRAS.339...12Z,2006MNRAS.368.1931L}. During the early phase, halos are formed with low concentration, and then the concentration increases during the second phase as the outer halo grows and the mass-accretion rate decreases. The concentration grows during this second phase because the virial radius setting the size of the whole halo increases while $r_s$ remains rather constant$^{\ref{this_footnote}}$. Halos of all masses undergo these two phases, but low mass halos complete the first phase early on and so they show large concentrations at present, whereas the very massive ones are still in the first phase. This process gives rise to the variation predicted by the numerical simulations shown in Fig.~\ref{fig:core_relation7}.  Contrary to the low mass halos,  the high mass halos show little evolution of the concentration with redshift (or, equivalently,  with time). According to this scenario, the actual $c$--$M_h$ relation should depend significantly on the cosmological parameters,  in particular, on $\sigma_8$ that parameterizes the amplitude of the matter density fluctuations in the early Universe, and on $\Omega_m$ that quantifies the total amount of matter.  The larger $\sigma_8$ or $\Omega_m$, the earlier the halos assemble and the larger the resulting concentration  \cite{2015MNRAS.452.1217C}.

%
%
%%%%%%
\subsection{Relation between DM core mass and stellar core radius}\label{sec:breakthrough}
The DM halo mass within the visible stellar core is
\begin{equation}
  M_{hc}= \frac{4\pi}{3}\rho_c r_{\star c}^3=\frac{4\pi g\kappa_c}{3}\,r_{\star c}^2,
 \label{eq:breakthrough1}
\end{equation}
with $\kappa_c$ the constant  $\rho_c r_c$, $r_{\star c}$ the stellar core radius,  and $g=r_{\star c}/r_c$. Provided $g\lesssim 1$, Eq.~(\ref{eq:breakthrough1}) gives the DM mass within the observed stellar core. Even if this is a relationship between the {\em core} DM halo mass and the stellar radius, it is encouraging to note that a similar relation is observed to hold between the DM core mass and the DM core radius (Fig.~\ref{fig:core_relation6b}), and between the {\em total} DM halo mass and the core radius; see the dashed line in Fig.~\ref{fig:core_relation6d}, corresponding to $M_h\propto r_c^{2.5}$. The baryon fraction in the core, defined as
\begin{equation}
  f_{bc}=   \frac{M_{\star c}}{M_{hc}}= \frac{\rho_{\star c} r_{\star c}}{g\kappa_c},
 \label{eq:breakthrough2}
\end{equation}
can be inferred  from the observed stellar mass surface density,  $\rho_{\star c} r_{\star c}$,   provided $g$ can be measured or estimated. 
Thus, if Eq.~(\ref{eq:maineq}) holds, from the stellar distribution alone one can estimate the DM core mass and the baryon fraction in the core. Using $\kappa_c$ from Eq.~(\ref{eq:obs_rhor}), Eqs.~(\ref{eq:breakthrough1}) and (\ref{eq:breakthrough2}) become,
\begin{equation}
M_{hc} \simeq 1.7^{+1.7}_{-0.8}\times 10^{5}\,M_\odot \left(\frac{r_{\star c}}{30\,{\rm pc}}\right)^2\,g,
\label{eq:breakthrough1b}
\end{equation}
and
\begin{equation}
  f_{bc}\simeq 2.2^{+2.1}_{-1.1}\times 10^{-3}\,\frac{\rho_{\star c} r_{\star c}}{0.1\,M_\odot\,{\rm pc}^{-2}}\,g^{-1},
 \label{eq:breakthrough2b}
\end{equation}
respectively. The error bars just consider the scatter in $\kappa_c$.

\begin{figure}
\centering 
\includegraphics[width=0.6\linewidth]{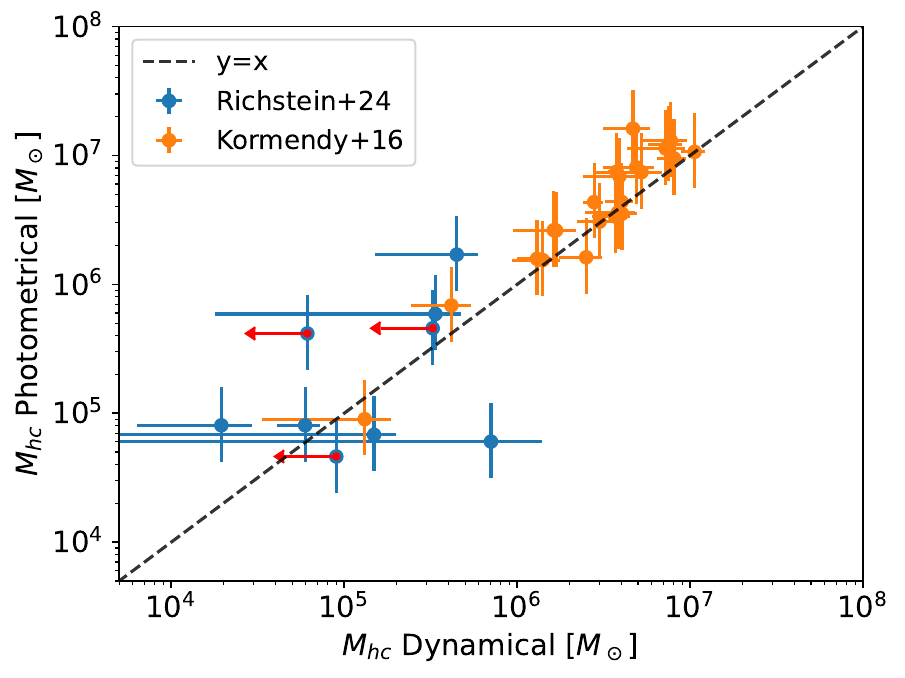}
\caption{
Comparison between the DM halo mass in the core of galaxies computed from the stellar velocity dispersion (horizontal axis) and from photometry alone as described by Eq.~(\ref{eq:breakthrough1b}) (vertical axis). The represented points include UFDs from Richstein+24~\cite{2024ApJ...967...72R} and dSphs from Kormendy+16~\cite{2016ApJ...817...84K}. The vertical error bars represent the dispersion in $\rho_c r_c$ (Eq.~[\ref{eq:obs_rhor}]) whereas the horizontal error bars account for the uncertainties in $\sigma_{\star c}$, as quoted in the original references.  The one-to-one line is shown as a dashed black line. The red arrows point out upper limits in the dynamical DM halo masses. 
}
\label{fig:core_relation10}
\end{figure}

In order to test the reliability of the above equations, we have used existing observations of ultra faint dwarfs (UFDs) and dwarf spheroidal galaxies (dSph)  to compare for individual galaxies the values of $M_{hc}$ computed from velocities and from Eq.~(\ref{eq:breakthrough1b}). The dynamical mass of a galaxy within $r_{\star c}$ can be computed from the observed velocity dispersion within the core radius,  $\sigma_{\star c}$, as
\begin{equation}
  M_{dyn} = \frac{2\ln 2}{G}\sigma_{\star c}^2\,r_{\star c},
  \label{eq:dyn_mass}
\end{equation}
with $G$ the gravitational constant.  In DM dominated systems,
\begin{equation}
  M_{hc}\simeq M_{dyn}.
  \label{eq:dyn_mass2}
\end{equation}
Equation~(\ref{eq:dyn_mass}) uses the definition in Eq.~(\ref{eq:core_def}) and assumes spherical symmetry as detailed by, e.g., \cite{2016ApJ...817...84K}. It differs from similar expressions found in the literature by factors of the order of one \cite{2024ApJ...967...72R}.
Figure~\ref{fig:core_relation10} shows the DM halo mass estimated from photometry (Eq.~[\ref{eq:breakthrough1b}]) versus
the value estimated from velocity dispersion (Eqs.~[\ref{eq:dyn_mass}] and [\ref{eq:dyn_mass2}]). The agreement is quite remarkable; often within the error bars set by Eq.~(\ref{eq:obs_rhor}). The UFDs have been included to show that the approximation works even in this extremely low mass regime, keeping in mind that part of the observed scatter away from the one-to-one relation is due to uncertainties in their dynamical mass estimate. The dynamical masses of UFDs are particularly uncertain because they are affected by the presence of stellar binaries, which may contribute to the velocity dispersion as much as the gravitational potential \cite[e.g.,][]{2022ApJ...939....3P}. The horizontal error bars in Fig.~\ref{fig:core_relation10} result from the statistical errors in $\sigma_{\star c}$, which are probably underestimating the real ones  since the effect of binaries is not included.   We have used $g=1$ for simplicity but the assumption $g\sim 1$ seems to be quite realistic \cite{2024ApJ...973L..15S,SanchezAlmeida24} and, eventually, it could be relaxed and refined if needed. 

Given the good agreement between the dynamical DM mass and the photometric DM mass represented in Fig.~\ref{fig:core_relation10}, Eq.~(\ref{eq:breakthrough1b}) seems to be a new valuable tool for estimating the DM halo mass from photometry alone. Photometry is much cheaper observationally than the spectroscopy required to determine the dynamical mass. The validity of Eq.~(\ref{eq:breakthrough1b})  implies  the validity of Eq.~(\ref{eq:breakthrough2b}), which also provides a new empirical way of estimating the baryon fraction in galaxies only from stellar photometry. Moreover, it tells us that the surface density of stars is a proxy for the baryon fraction in the inner parts of a galaxy.

The above estimate can be extended to the mass of the whole DM halo using a model to represent the DM halo beyond the core (e.g., the piecewise profile in Eq.~[\ref{eq:cases}] and Fig.~\ref{fig:core_relation4}).  Thus,  $M_{hc}$ can be used to estimate $M_h$. To have a first idea of the ratio between them, assume that the stelar core radius is not very different from the matching radius $r_m$ that separates the inner and outer parts of the piecewise profile (Fig.~\ref{fig:core_relation4}), which is a quite common assumption in the literature \cite[e.g.,][]{2016JCAP...03..009L,2023MNRAS.523.4786O}. Then, the ratio of masses turns out to be
\begin{equation}
  M_h/M_{hc}\simeq \frac{\ln(1+r_{\star c}/r_s)-(r_{\star c}/r_s)/(1+r_{\star c}/r_s)}{\ln(1+c)-c/(1+c)},
\end{equation}
which varies from a few to a factor of ten when the concentration varies as predicted, from $c\sim5$ in high mass halos to $c\sim 20$ in low mass halos (Fig.~\ref{fig:core_relation7}, the blue lines).  % see core_relation11.py
%
%%%%%%
\subsection{Constant DM dynamical pressure}

The dynamical pressure in a fluid scales like the density times the square of the characteristic velocity. Thus, for the DM in the core, the effective DM dynamical pressure is 
\begin{equation}
  P_c\propto \rho_c \sigma_c^2,
\end{equation}
with $\sigma_c$ the velocity dispersion of the DM particles in the core. Assuming the DM cores to be virialized (i.e., assuming that Eqs.~[\ref{eq:dyn_mass}] and [\ref{eq:dyn_mass2}] hold for the DM particles too), then 
\begin{equation}
  P_c\propto (\rho_c\,r_c)^2 \simeq {\rm constant},
 \label{eq:const_press}
\end{equation}
so that Eq.~(\ref{eq:maineq}) implies that the dynamical pressure to be exerted by the DM particles if they could collide would be the same in all halos, independently of their total mass or size.   But collisionless CDM particles do not collide, and Eq.~(\ref{eq:const_press}) has to be interpreted as a property that emerges from the existence of the $c$--$M_h$ relation. 

%
%%%%%
%
%\section{Discussion}
%
%
\section{Conclusions}\label{sec:conclusions}

We reviewed the observational evidence for $\rho_c r_c\simeq$ constant (Eq.~[\ref{eq:maineq}]; Sect.~\ref{sec:observations}) and then put forward a simple version of the commonly accepted interpretation behind it (Sect.~\ref{sec:theory}). Equation~(\ref{eq:maineq}) requires the existence of a core in the DM distribution. Halos formed in DM-only CDM cosmological numerical simulations do not have inner cores but cusps (Eq.~[\ref{eq:nfw}]), however, if any physical process redistributes the DM particles of the expected CDM halos then Eq.~(\ref{eq:maineq}) is satisfied automatically. It emerges from the relation between concentration and DM halo mass expected in $\Lambda$CDM cosmological simulations. This relation is set by the time of halo formation, so that low mass halos formed earlier and now they present larger concentration (Sect.~\ref{sec:what_sets}). The conventional explanation to understand how the original cuspy CDM halos become cored halos is {\em stellar feedback}. This term encapsulates all the baryon driven processes that shuffles gas and mass around (e.g., supernova explosions or stellar winds), modifying  the overall potential, including the distribution of DM particles in the center of galaxies \cite{2010Natur.463..203G,2014Natur.506..171P}. However, this transformation is not specific to stellar feedback, keeping in mind that any physical process that thermalizes a self gravitating structure tends to form cores \cite{1993PhLA..174..384P,2020A&A...642L..14S}. Thus, any other sensible physical process that redistributes matter without altering the original mass of the CDM halos is able to account for Eq.~(\ref{eq:maineq}). In other words, the property of $\rho_cr_c$ to be approximately constant is not specific to CDM but, rather, it is also  expected in many alternative DM theories forming cores \cite[e.g.,][]{2016JCAP...03..009L,2016PhRvL.116d1302K,2022MNRAS.517.3045C}. Theories that only redistribute mass to produce cores have the advantage of leaving the large scale structure of the Universe unchanged, thus being in  agreement with the standard $\Lambda$CDM.

The mathematical development in Sect.~\ref{sec:theory}  parallels others existing in the literature, except that the core is modeled with a  different expression \cite[e.g.,][]{2016JCAP...03..009L,2024PASJ...76.1026K}. Here we provide a full account of the derivation of the main equations for the sake of comprehensiveness, which help us to make the qualitative comparison with observations  in Sect.~\ref{sec:theory_observations}. However,  we could have started off by assuming the relevant Eqs.~(\ref{eq:constraint1}) and (\ref{eq:constraint2}) and proceed from here. This loose dependence of the results on the actual shape of the core is consistent with the fact that other alternative forms of the piecewise profile with core that we tried (top hat profiles) render qualitatively similar results.

The agreement between the simple theory and observations is notable, keeping in mind that there is no fitting or fine tuning in matching lines and points  in Fig.~\ref{fig:core_relation9}. Even more, the theory predicts a moderate increase of $\rho_cr_c$ with $M_h$ similar to the one hinted at by the observations. However, the best fitting $c$--$M_h$ relations correspond to large cores (the green dashed line represents $r_m/r_s=1.8$) or $z\not= 0$ (the solid orange and green lines in Fig.~\ref{fig:core_relation9} correspond to $z=1$). The latter is a result that we do not understand; even if the transformation of cusps to cores requires time and starts at high redshift, the accretion of DM in the outskirts of the halos should  continue all the way to the present, a process leading to the $c$--$M_h$ relation at $z=0$. As we discuss in Sect.~\ref{sec:what_sets}, the $c$--$M_h$ depends on the cosmological parameters $\sigma_8$ and $\Omega_m$ since they set the assembly time of the DM halos. Varying them may improve the agreement when employing the theoretical $c$--$M_h$ relations at $z=0$,  but we have not pursued this idea further.  

As a byproduct of the effort to compile $\rho_c r_c$ values, we show that the fact that the product is constant can be  used to estimate the mass in the DM halo of a galaxy from the distribution of stars alone. This possibility can be very useful for low stellar mass galaxies where the determination of their DM content using traditional kinematical measurements is technically difficult, whereas their photometry is doable.  The same argument allows one to estimate the baryon fraction in the core of these systems. Dwarf galaxies also tend to show a core in the {\em stellar} distribution \cite[e.g.,][]{2021ApJ...922..267C,2024A&A...681A..15M}, with the radii of the stellar and DM cores expected to scale with each other \cite{2024ApJ...973L..15S,SanchezAlmeida24}. This idea plus Eq.~(\ref{eq:obs_rhor})  allow us to propose specific relations between the observed stellar core radius and the DM core mass  (Eq.~[\ref{eq:breakthrough1b}]) and between the observed stellar mass surface density and the baryon fraction in the core  (Eq.~[\ref{eq:breakthrough2b}]). The latter tells us that the surface density of stars is a proxy for the baryon fraction in the inner parts of a galaxy.
The proposed calibrations are in good agreement with DM masses estimated from dynamical measurements in low mass galaxies (Fig.~\ref{fig:core_relation10}). Note that the numerical coefficients of the proposed scaling laws depend on the definition of core radius, for which we adopted Eq.~(\ref{eq:core_def}). Other definitions can be trivially recalibrated.

%
%%%%%%%
%%%%%%%%%%%%%%%%%%%%%%%%%%%%%%%%%%%%%%%%%%
%%%%%%%%%%%%%%%%%%%%%%%%%%%%%%%%%%%%%%%%%%
\vspace{6pt}

\funding{
This research has been partly funded through grant PID2022-136598NB-C31 (ESTALLIDOS8) by MCIN/AEI/10.13039/501100011033 and by “ERDF A way of making Europe”. It was also supported by the European Union through the grant  ''UNDARK'' of the widening participation and spreading excellence programme (project number 101159929).
}

\dataavailability{All the data used in this paper are publicly available in the cited references.}

% Only for journal Nursing Reports
%\publicinvolvement{Please describe how the public (patients, consumers, carers) were involved in the research. Consider reporting against the GRIPP2 (Guidance for Reporting Involvement of Patients and the Public) checklist. If the public were not involved in any aspect of the research add: ``No public involvement in any aspect of this research''.}

% Only for journal Nursing Reports
%\guidelinesstandards{Please add a statement indicating which reporting guideline was used when drafting the report. For example, ``This manuscript was drafted against the XXX (the full name of reporting guidelines and citation) for XXX (type of research) research''. A complete list of reporting guidelines can be accessed via the equator network: \url{https://www.equator-network.org/}.}

% Only for journal Nursing Reports
%\useofartificialintelligence{Please describe in detail any and all uses of artificial intelligence (AI) or AI-assisted tools used in the preparation of the manuscript. This may include, but is not limited to, language translation, language editing and grammar, or generating text. Alternatively, please state that “AI or AI-assisted tools were not used in drafting any aspect of this manuscript”.}

\acknowledgments{
I am grateful to Ignacio Trujillo for bringing to my attention the empirical relationship explored in the work (Eq.~[\ref{eq:maineq}]). I am also thankful to him, Claudio Dalla Vecchia, Angel Ricardo Plastino, Camila Correa, and Andrés Balaguera for enlightening discussions and clarifications on various issues addressed in the manuscript. 
}

\conflictsofinterest{The author declares no conflicts of interest.} 

%%%%%%%%%%%%%%%%%%%%%%%%%%%%%%%%%%%%%%%%%%
%% Optional

%% Only for journal Encyclopedia
%\entrylink{The Link to this entry published on the encyclopedia platform.}

\abbreviations{Abbreviations}{
The following abbreviations are used in this manuscript:\\

\noindent 
\begin{tabular}{@{}ll}
  DM & Dark matter\\
  CDM & Cold dark matter\\
  dSph & Dwarf spheroidal galaxy\\
  $\Lambda$CDM & Concordance cosmological model\\
  UFD & Ultra faint dwarf\\
\end{tabular}
}

%%%%%%%%%%%%%%%%%%%%%%%%%%%%%%%%%%%%%%%%%%
%% Optional
\appendixtitles{no} % Leave argument "no" if all appendix headings stay EMPTY (then no dot is printed after "Appendix A"). If the appendix sections contain a heading then change the argument to "yes".
\appendixstart
\appendix
\section[\appendixname~\thesection]{Bibliography on the $\rho_c r_c$ versus $M_h$ relation}\label{app:literature}
\begin{figure}
\centering 
\includegraphics[width=0.8\linewidth]{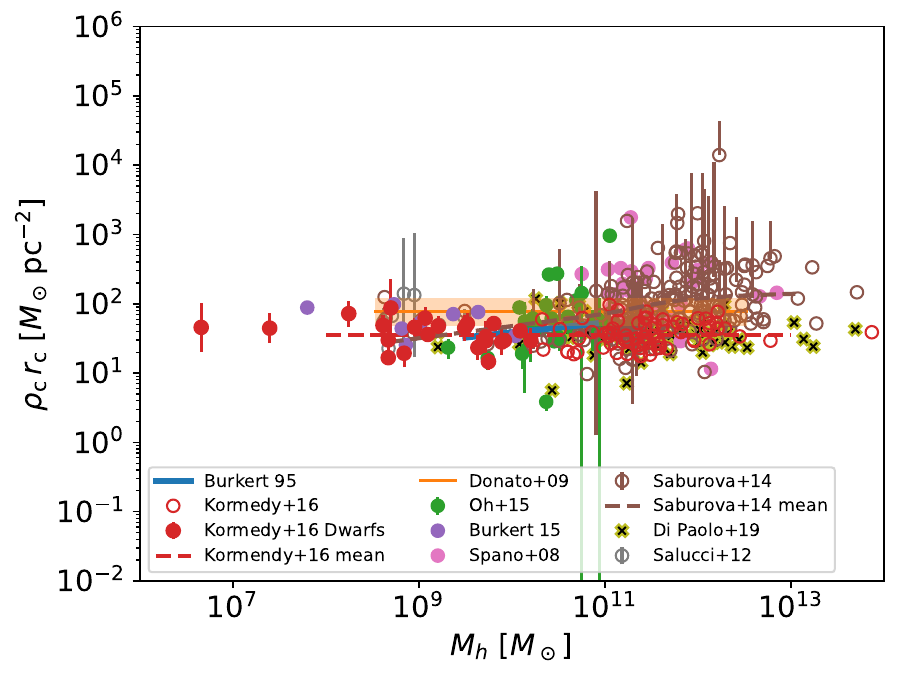}
\caption{
Figure identical to  Fig.~\ref{fig:core_relation6}  except that the range of the ordinates ($\rho_c r_c$) has been expanded to show the same eight orders of magnitude variation as the  DM halo mass range ($M_h$). For the rest of details, see Fig.~\ref{fig:core_relation6}.
}
\label{fig:core_relation6_a}
\end{figure}

\begin{table}[H] 
\caption{References used to constraint $\rho_cr_c$.\label{tab:table1}}
%\newcolumntype{C}{>{\centering\arraybackslash}X}
\begin{tabularx}{\textwidth}{LCCL}
\toprule
\textbf{Reference}	& \textbf{$\rho_cr_c$ $[M_\odot\,{\rm pc}^{-2}]$\,\textsuperscript{1}}	& \textbf{$\log M_h$ $[M_\odot]$\,\textsuperscript{2}} & \textbf{Comment}\,\textsuperscript{3}\\
\midrule
\cite{1995ApJ...447L..25B}~\citeauthor{1995ApJ...447L..25B}~(\citeyear{1995ApJ...447L..25B})  & $41.5\pm 5.9$ & $10.2\pm 0.4$ & Corrected $r_c$ \& $M_h$\\
\cite{2009MNRAS.397.1169D}~\citeauthor{2009MNRAS.397.1169D}~(\citeyear{2009MNRAS.397.1169D})& $76^{+43}_{-16}$&	$8.5$\,--\,$12.5$& Corrected $r_c$; $M_h$ from $M_B$ \\
\cite{2015ApJ...808..158B}~\citeauthor{2015ApJ...808..158B}~(\citeyear{2015ApJ...808..158B})&$64^{+56}_{-34}$ &$9.0\pm 0.6$& Corrected $r_c$\\
\cite{2015AJ....149..180O}~\citeauthor{2015AJ....149..180O}~(\citeyear{2015AJ....149..180O})&$67\pm 65$ &$10.4\pm 0.4$& sigma-clipping in noise\\
\cite{2016ApJ...817...84K}~\citeauthor{2016ApJ...817...84K}~(\citeyear{2016ApJ...817...84K}) &$39 \pm 17$&$11.5 \pm 0.6$&Massive galaxies. Corrected $r_c$\\
\cite{2016ApJ...817...84K}~\citeauthor{2016ApJ...817...84K}~(\citeyear{2016ApJ...817...84K})&$40 \pm 17$&$9.1 \pm 0.8$&Dwarfs. Corrected $r_c$\\
\cite{2016ApJ...817...84K}~\citeauthor{2008MNRAS.383..297S}~(\citeyear{2008MNRAS.383..297S})&$230 \pm 300$&$11.5 \pm 0.5$&Corrected $r_c$\\
\cite{2014MNRAS.445.3512S}~\citeauthor{2014MNRAS.445.3512S}~(\citeyear{2014MNRAS.445.3512S})&$59 \pm 36 $&$8.6$\,--\,$11$& Only low mass. Corrected $r_c$.\\
\cite{2012MNRAS.420.2034S}~\citeauthor{2012MNRAS.420.2034S}~\citeyear{2012MNRAS.420.2034S}&$71\pm 40$&$9.0\pm 0.4$&dSph only. Corrected $r_c$\\
\cite{2019MNRAS.490.5451D}~\citeauthor{2019MNRAS.490.5451D}~\citeyear{2019MNRAS.490.5451D}&$41\pm 21$&$9.2$\,--\,$13.7$&Corrected $r_c$, using their $M_h$.\\
\bottomrule
\end{tabularx}
\noindent{\footnotesize{\textsuperscript{1} Mean and standard deviation of the values mentioned in the reference.}}

\noindent{\footnotesize{\textsuperscript{2} Mean and standard deviation or range of values.}}

\noindent{\footnotesize{\textsuperscript{3} Further details given in Appendix~\ref{app:literature}.}}
\end{table}

This appendix details the use of the bibliography leading to Figs.~\ref{fig:core_relation6_a}, \ref{fig:core_relation6},  \ref{fig:core_relation6f}, \ref{fig:core_relation6b}, \ref{fig:core_relation6d}, and \ref{fig:core_relation9}. Since the estimate of the parameters is cumbersome, we discuss the main issues and assumptions in this Appendix and in Table~\ref{tab:table1}. The various references are identified in the figures through the corresponding insets.  
\begin{itemize}
\item[-] \citeauthor{1995ApJ...447L..25B}~(\citeyear{1995ApJ...447L..25B}) \cite{1995ApJ...447L..25B}  explicitly gives a relation between central density and core radius and between halo mass and core radius. Pieced together, they provide the relation represented in Fig.~\ref{fig:core_relation6} with $M_h$ within the range represented in his Fig.~3. The original relations have to be corrected to our core radius definition (Eq.~[\ref{eq:core_def}]) and to the total halo mass (his Eq.~[4]).
\item[-]  \citeauthor{2009MNRAS.397.1169D}~(\citeyear{2009MNRAS.397.1169D}) \cite{2009MNRAS.397.1169D}.  The value with error bars is directly given in the paper. They conclude that the product $\rho_cr_c$ is constant for absolute magnitudes $M_B$ from -7 to -22. In order to transform these $B$ magnitudes into halo masses, (1) we use a stellar mass to light ratio $M_\star/L_\star$ of one (in solar units) and then use $M_\star$ to estimate $M_h$ using the halo to stellar mass ratio at redshift zero from \cite{2013ApJ...770...57B}. They use the same definition of core radius as \cite{1995ApJ...447L..25B}, and so has to be corrected to ours in Eq.~(\ref{eq:core_def}).
\item[-] \citeauthor{2015ApJ...808..158B}~(\citeyear{2015ApJ...808..158B})~\cite{2015ApJ...808..158B}. We take $\rho_c$ and $r_c$ from \cite{2015ApJ...808..158B}, and the corresponding $M_\star$ from \cite{2012AJ....144....4M}. Then $M_h$ was estimated using the halo to stellar mass ratio from \cite{2013ApJ...770...57B}. The conversion between the core radius used in the original work and Eq.~(\ref{eq:core_def}) was carried out based on Fig.~1 of \cite{2015ApJ...808..158B}.
\item[-] \citeauthor{2015AJ....149..180O}~(\citeyear{2015AJ....149..180O})~\cite{2015AJ....149..180O} do not determine the product $\rho_cr_c$, but they provide $\rho_c$ and $r_c$ separately. They also provide the absolute $V$ magnitude $M_V$ which, assuming a mass to light ratio of one, allows us to estimate $M_h$ using the DM halo to stellar mass ratio from \cite{2013ApJ...770...57B}. The $r_c$ used in this reference happens to agree with  Eq.~(\ref{eq:core_def}) and so we do not change it. The averages in Table~\ref{tab:table1} were computed after removing the  $\rho_c r_c$ values with larger error (see Fig.~\ref{fig:core_relation6}).
\item[-]  \citeauthor{2016ApJ...817...84K}~(\citeyear{2016ApJ...817...84K})~\cite{2016ApJ...817...84K} is the reference with the largest number of galaxies. It gives clear relations between $\rho_c$ and $r_c$, and $M_B$. The galaxies are separated in low and high masses. As for many of the above references, $M_h$ is obtained from their $M_B$ assuming a stellar mass to light ratio of one, and using the scaling between stellar and halo mass in \cite{2013ApJ...770...57B}. For the core radius, the authors directly provide the scaling between their core radius and  Eq.~(\ref{eq:core_def}). % Sect. 3. They have many different; before Eq. 4, r_c = 2**0.5/3 * their_radius ... but is not alwys the same. I will use 0.5 ...  
\item[-]\citeauthor{2008MNRAS.383..297S}~(\citeyear{2008MNRAS.383..297S})~\cite{2008MNRAS.383..297S} also find approximately constant $\rho_c r_c$. The galaxies are fairly massive (see Table~\ref{tab:table1}). No error bars are given. We transform their $r_s$ into ours.
\item[-] \citeauthor{2014MNRAS.445.3512S}~(\citeyear{2014MNRAS.445.3512S})~\cite{2014MNRAS.445.3512S} compile a  large list of objects from various sources. The authors compute and provide the product $\rho_c r_c$. We infer $M_h$ from $M_B$ as explained above. The points without error bars in Fig.~\ref{fig:core_relation6} are not points with zero error but points without an estimate of the error. They claim a variation with luminosity so that the more luminous (and so more massive) galaxies have larger $\rho_s r_s$ (see Fig.~\ref{fig:core_relation6}). The low mass value is consistent with other estimates. They use a Burkert DM halo to define the radius, which we transform to our definition in Eq.~(\ref{eq:core_def}). 
\item[-] \citeauthor{2012MNRAS.420.2034S}~(\citeyear{2012MNRAS.420.2034S})~\cite{2012MNRAS.420.2034S}. We consider only the data for the dwarf spheroidal galaxies (dSph).
\item[-] \citeauthor{2019MNRAS.490.5451D}~(\citeyear{2019MNRAS.490.5451D})~\cite{2019MNRAS.490.5451D}. These are low surface brightness  galaxies, but seem to behave as the rest. Galaxies are stacked in halo mass bins. We take the halo mass from them and then correct $r_c$ to accommodate their definition (Burkert profile) into our definition (Eq.~[\ref{eq:core_def}]). 
\end{itemize}

\section{The theoretical value of $\rho_c r_c$ when $r_m=r_s$}\label{app:appb}

In the case when the matching radius of the piecewise profile is equal to the characteristic radius of the corresponding NFW profile ($r_m/r_s=1$ in Fig.~\ref{fig:core_relation4}) then  several numerical coincidences happen and  $\rho_c r_c$ and $\rho_s r_s$ are almost equal,
\begin{equation}
  \frac{\rho_cr_c}{\rho_s r_s}=\frac{8[3(\ln 2-1/2)]^{5/2}\,[2^{2/5}-1]^{1/2}}{[12(\ln 2-1/2)-1]^{1/2}}\simeq 1.0068\dots.
\end{equation}
It follows from Eqs.~(\ref{eq:silly}), (\ref{eq:constraint1}), and  (\ref{eq:constraint2}) when  $r_m=r_s$.

 %

%%%%%%%%%%%%%%%%%%%%%%%%%%%%%%%%%%%%%%%%%%
\begin{adjustwidth}{-\extralength}{0cm}

\reftitle{References}


%\bibliography{your_external_BibTeX_file}
\newcommand{\prd}{Physical Review D}
\newcommand{\prl}{Phys. Rev. Lett.}
\newcommand{\apjl}{ApJL}
\newcommand{\aj}{AJ}
\newcommand{\aap}{A\&A}
\newcommand{\aapr}{A\&ARev}
\newcommand{\aaps}{A\&AS}
\newcommand{\apj}{ApJ}
\newcommand\apjs{ApJS}
\newcommand{\mnras}{MNRAS}
\newcommand{\pasa}{PASA}
\newcommand{\jcap}{JCAP}
\newcommand{\nat}{Nat}
\newcommand{\ssr}{SSRv}
\newcommand{\pasj}{PASJ}
%\bibliography{../figures+/biblio}
%\bibliography{biblio}

% If authors have biography, please use the format below
%\section*{Short Biography of Authors}
%\bio
%{\raisebox{-0.35cm}{\includegraphics[width=3.5cm,height=5.3cm,clip,keepaspectratio]{Definitions/author1.pdf}}}
%{\textbf{Firstname Lastname} Biography of first author}
%
%\bio
%{\raisebox{-0.35cm}{\includegraphics[width=3.5cm,height=5.3cm,clip,keepaspectratio]{Definitions/author2.jpg}}}
%{\textbf{Firstname Lastname} Biography of second author}

% For the MDPI journals use author-date citation, please follow the formatting guidelines on http://www.mdpi.com/authors/references
% To cite two works by the same author: \citeauthor{ref-journal-1a} (\citeyear{ref-journal-1a}, \citeyear{ref-journal-1b}). This produces: Whittaker (1967, 1975)
% To cite two works by the same author with specific pages: \citeauthor{ref-journal-3a} (\citeyear{ref-journal-3a}, p. 328; \citeyear{ref-journal-3b}, p.475). This produces: Wong (1999, p. 328; 2000, p. 475)

%%%%%%%%%%%%%%%%%%%%%%%%%%%%%%%%%%%%%%%%%%
%% for journal Sci
%\reviewreports{\\
%Reviewer 1 comments and authors’ response\\
%Reviewer 2 comments and authors’ response\\
%Reviewer 3 comments and authors’ response
%}
%%%%%%%%%%%%%%%%%%%%%%%%%%%%%%%%%%%%%%%%%%
\PublishersNote{}
\end{adjustwidth}
\end{document}